# Instrumentation Status of the Low-β Magnet Systems at the Large Hadron Collider (LHC)

* Darve C., Balle C., Casas-Cubillos J., Perin A., Vauthier N.

* Fermi National Accelerator Laboratory, PO. Box 510, Batavia, IL, 60510, USA
European Organization for Nuclear Research, TE department, Geneva, 1211, CH

The low-β magnet systems are located in the Large Hadron Collider (LHC) insertion regions around the four interaction points. They are the key elements in the beams focusing/defocusing process allowing proton collisions at luminosity up to $10^{34} cm^{-2} s^{-1}$. Those systems are a contribution of the US-LHC Accelerator project. The systems are mainly composed of the quadrupole magnets (triplets), the separation dipoles and their respective electrical feed-boxes (DFBX). The low-β magnet systems operate in an environment of extreme radiation, high gradient magnetic field and high heat load to the cryogenic system due to the beam dynamic effect. Due to the severe environment, the robustness of the diagnostics is primordial for the operation of the triplets. The hardware commissioning phase of the LHC was completed in February 2010. In the sake of a safer and more user-friendly operation, several consolidations and instrumentation modifications were implemented during this commissioning phase. This paper presents the instrumentation used to optimize the engineering process and operation of the final focusing/defocusing quadrupole magnets for the first years of operation.

INTRODUCTION

The low-β magnet systems are composed of four quadrupole magnets (Q1, Q2a, Q2b, Q3) collectively called the inner triplet, a beam separation dipole magnet (D1), corrector magnets and the electrical feed-boxes, (DFBX). The DFBX is used to operate the inner triplet magnets, using 7.5 kA, 600 A and 120 A current leads. The D1 is superconducting at the low luminosity IPs (IP2 and IP8) and resistive at the high luminosity IPs. These magnets provide the final focusing/defocusing of the proton beams before/after collision. Figure 1 shows the layout of the different magnets and of the DFBX.
   Several non-conformities were discovered during the hardware and beam commissioning of the LHC [1-2]. The status of the low-β magnet system instrumentation is reported.

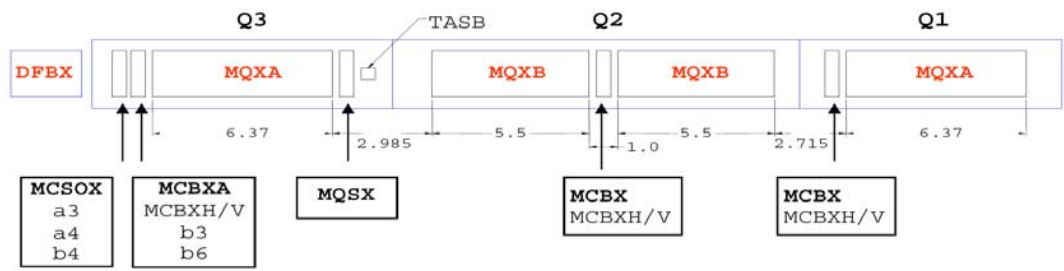

Figure 1 Layout of the low-β magnet system

SYSTEM DESCRIPTION

Type of instrumentation
Equipment TAG code system as being use [3]. We use this terminology, hereafter
  The DFBX instrumentation is composed of:
- TT8xx: RTDs (Pt100, Cernox$^{TM}$)
- PT8xx: pressure transducers based on passive strain gauges
- LT8xx: liquid helium level gauges based on superconducting wire
- EH8xx: cryo-electrical heaters
- CV8xx: control valve

  The only temperature sensors, which are not redundant, are the Pt100 installed during the consolidation of the vapor cool lead cooling [1].
Voltage taps and quench protection heaters are used to monitor and to protect the current leads and the inner triplet magnets.
  Control valves regulate the cooling of the 7.5 kA HTS leads and of the 600 A and 120 A vapor cooled current leads, which are entirely resistive (VCL). Several control valves are also used to regulate the level of liquid helium in the DFBX.
  The type of sensors used for the inner triplets are identical to the ones used in the arcs of the LHC. Sensors are immersed in the pressurized superfluid helium bath or are installed on the cold masses in the insulation vacuum [4]. Figure 2 shows the synoptic of the low-β magnet system in L2.

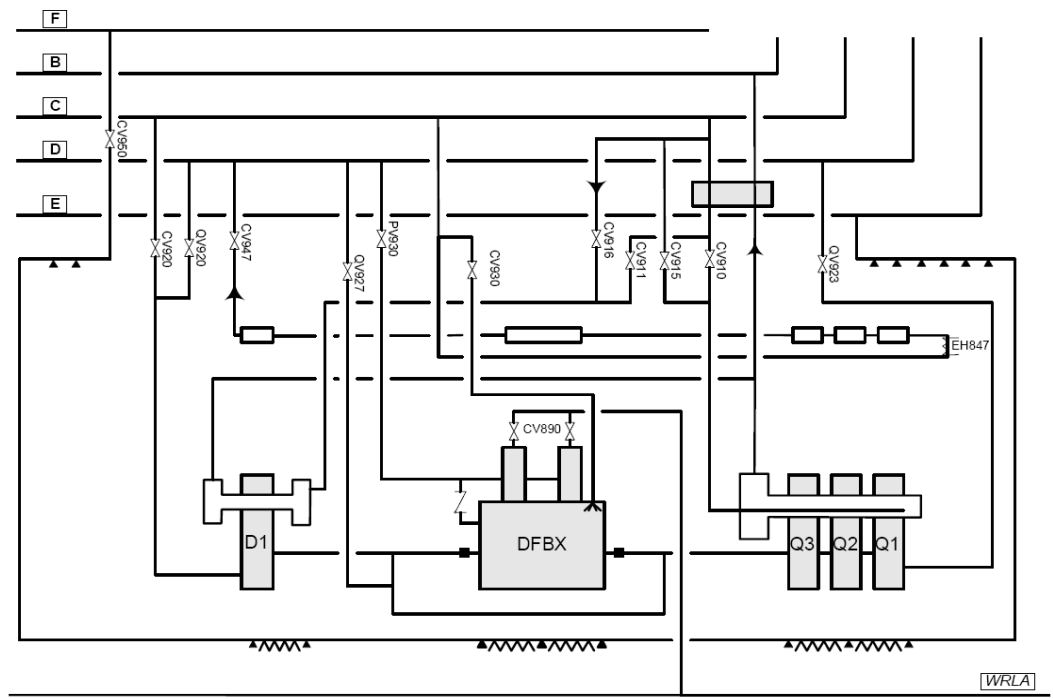

Figure 2 Synoptic of the low-β magnet system

  Although no measurement at sub-atmospheric conditions was performed before 2010, a helium guard was added in prevision of saturated helium pressure measurement of the inner triplet bayonet heat exchanger.
  In addition to the cryogenic instrumentation, the triplets are equipped with several other sensors and control devices, in particular for the precise alignment of the magnets [5].

Diagnostic and control system

The monitoring and control of the low-β magnet system instrumentation are transmitted from the measuring elements to the Front End Computer (FEC). Since the low-β magnet systems are part of the LHC long straight sections, the control also performed by using Programmable Logic Controller (PLC). The control system is distributed and usesWorldFIP for the RadTol equipment and Profibus for commercial IOs and intelligent valve positioners [6]. The electronics is installed in protected area because of the high radiation dose that is expected inside the tunnel in the inner triplets area.

Following the LHC commissioning, the control logic was simplified and the interlocks were modified. The low-β magnet system diagnostic uses the same tools than the rest of the LHC machine.

Flow measurement for the resistive current leads cooling and other control valves

The resistive current leads of the DFBXs were supposed to be operated by controlling their mass flow as a function of the electrical current. However, due to the difficulty to develop radiation resistant flowmeters and in order to limit the instrumentation cost and complexity, no flowmeter was installed on the triplet systems. An attempt was made to use the control valves as orifice flowmeters, but the difficulty to get a reliable reading of the position of the control valves did not allow to reach a precision usable for controlling the mass flow with the required precision. A technique using the temperature near the room temperature head of the current leads as a way to control the mass flow was developed and successfully implemented during the hardware commissioning. However, the possibility to implement 1) a remote flow measurement reading or 2) a nuclear qualified flow transmitter, is envisaged to increase the quality a reliability of the regulation.

Documentation and traceability – MTF

Every cryogenic measuring element shall be recorded in the CERN Manufacturing and Test Folder (MTF) [7]. Equipments are defined in the database and are linked in an assembly breakdown structure. For instance, Figure 3 shows the first level of the equipment and instrumentation located on the DFBX top plate, including the four vapor cooled current leads. The second level lists the temperature sensors, valves and heaters on the given vapor cooled lead chimney. Individual equipment data is available via MTF. Non-conformity, electrical quality assurance and radiation survey can be also tracked using the MTF database.

Figure 3: MTF structure for the DFBXA. To be improved

## INSTRUMENTATION PERFORMANCE

### Performance under physical environment

In addition to the use of instrumentation and signal conditioners rated for cryogenic and/or vacuum applications, the main physical constraints to the operation of the low-β magnet systems are:

- Magnetic field: The quadrupole magnets are operated in superfluid helium at 1.9 K with a nominal gradient of 205 T/m in the 70 mm bore [8]. Voltage taps are installed and temperature sensors are installed at the proximity of the bore.
- Radiation: The environment of the low-β magnet system will be exposed to a complex radiation distribution pattern due to the proximity of the interaction points [9]. For comparison, the expected nominal LHC operation annual radiation dose for the arc magnet and for the CMS/ATLAS low-β regions are 1 and 1000 Gy, respectively. The beam induced radioactivity of the materials will increase the complexity of interventions after prolonged operation.

The LHC instrumentation was specifically validated to ensure a reliable operation in the radiation environment [10-14]. Irradiation campaigns have been performed at CERN for pressure sensors and for all tunnel electronics, albeit for doses that corresponded to the LHC dispersion-suppressor areas. The doses at the IT will be much higher, however the sensing instrumentation is not expected to change significantly for higher doses. The instrumentation installed on the low-β magnet systems can withstand a large radiation dose and a high magnetic field.

According to the vendor, Cernox$^{TM}$ temperature sensors show only a -10 mK radiation induced offset at 4.2 K when exposed to a gamma integrated radiation dose of $10^4$ Gy. The cryogenic thermometers have been qualified in nominal conditions (1.8 K) to radiation doses up to $4 \cdot 10^{14}$ n/cm$^2$ [10] and apparently the temperature drift with radiation saturates to about 1mK. In nominal conditions the radiation to which the cold mass will be exposed is $2.10^9$ n.cm$^{-2}$.s$^{-1}$, resulting in a negligible 0.2 mK temperature increase assuming a beam heating of $2.10^{-10}$ mK/(n.cm$^{-2}$.s$^{-1}$).

The choice of the control valve shall also comply with the high radiation dose, they have a split design with the pneumatics part installed inside the tunnel and the control electronics installed in radiation protected areas. The requested radiation tolerance for the electronics is 200 Gy and this equipment was never tested to the high ration levels that will be found in the ITs.

The choice of materials was adapted to the physical environment, e.g. PEEK insulator blocks are used for the electrical connectors [15].

### Temperature measurement stability

The temperature sensor requirements requested an overall accuracy of +/- 10 mK; however it is expected that they must remain better than 0.25% (5 mK at 2 K) during the whole machine lifetime.

All critical sensors are either redundant or exchangeable without breaking the insulation vacuum.

The precision of the inner triplet superfluid temperature measurement was qualitatively estimated by using the "lambda transition". For a given pressure, the transition of the superfluid helium from its He I phase to its He II phase allows to calibrate the Cernox$^{TM}$ RTD immersed in the superfluid helium. Figure 4 shows the cool-down temperature of four sensors installed along the inner triplet. The lambda transition occurs at a temperature of 2.17 K for a pressure of 1.3 bar. The standard deviation is 0.343 K, and a maximum offset of 13 mK was measured along the 40 m long inner triplet cold mass string. This validation was performed on all triplets to detect faults in the sensors or in the acquisition chain. However temperature homogeneity is a much better measurement condition to qualify the measurement chain and to evaluate the dispersion between the different sensors.
.

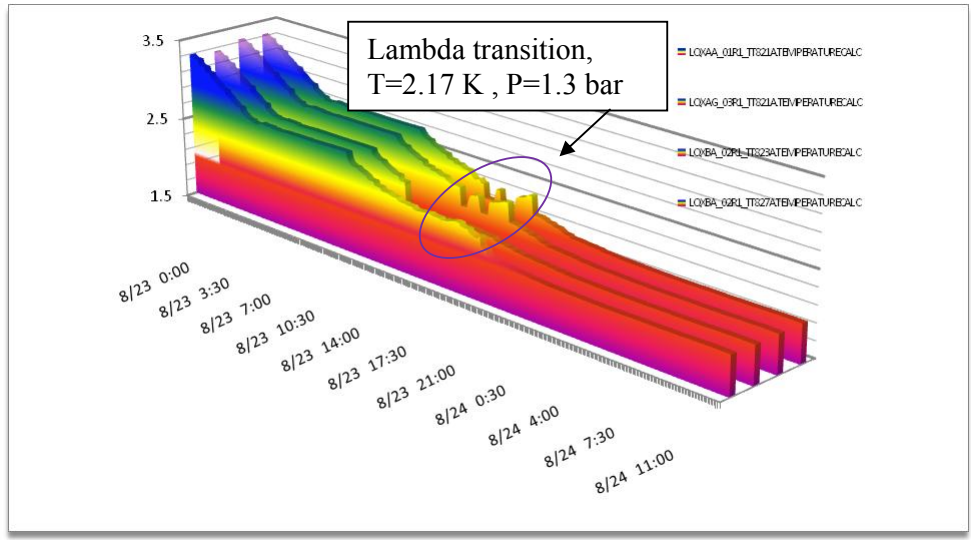

Figure 4: Lambda transition for four temperature sensors located in the inner triplet.

Identified and corrected non-conformities

The different consolidations completed on the DFBX are detailed elsewhere [CEC09]. After the first year of LHC operation, we can report that the following consolidations have been validated:

- The stable cooling and powering of the VCL were possible after a complete installation of new VCL temperature sensors. The performance of the new control system was successfully tested.
- The low-β magnet system was tested for a ramp and acceleration rates as larger as 6 A/s and 0.4 A/s$^2$, respectively. This performance is sufficient for the first year of LHC operation.
- Passive heaters composed of copper braids installed on the helium collector of the low-β magnet sub-atmospheric system, are usable.
- The electronic card of the 25 ohm resistance warm-up heaters permit to heat each of the quadrupole cold mass up to 100 W.
- Electronic filters were installed in order to prevent temperatures offset, which were observed during field measurement performed on the newly installed temperatures sensors located on the 600 A VCL cooling systems. These offsets were the result of electromagnetic noise induced onto the measurement chain.
- Temperature switches were installed on the safety relief valve outlet, giving the possibility to monitor a possible helium leak, in particular after cold helium release.

The safe operation of the low-β magnet system was successfully tested using the new implemented equipment and interlocks [15].

CONCLUSION

The performance of the instrumentation and cryo-equipment used for the low-β magnet system is conformed to expectation. The radiation environment has a limited impact of the measurement. The temperature sensors precision remains better than 0.25% (5mK at 2K) during the whole machine lifetime. Development of the instrumentation traceability was possible using MTF.

The commissioning and operation of the low-β magnet system have permitted to validate consolidation actions and the new instrumentation. The low-β magnet system equipment is now safe to operate and complies with the LHC requirements.

ACKNOWLEDGEMENT

The authors would like to thanks the TE/CRG personnel for their technical support during the LHC hardware commissioning, including the cryogenic operation section. The testing of the low-β magnet system was possible thanks to the hard work of the hardware commissioning and operation team.